\documentclass[
    ,final            
  ]
  {aipproc}

\layoutstyle{8x11single}
\usepackage{sidecap}

\begin{document}

\title{A unitary and causal effective field theory}


\author{A.M.  Gasparyan}{
  address={GSI Helmholtzzentrum f\"ur Schwerionenforschung GmbH, Planckstrasse 1, 64291 Darmstadt, Germany}
,altaddress={SSC RF ITEP, Bolshaya Cheremushkinskaya 25, 117218 Moscow,
 Russia}
}

\author{M.F.M. Lutz}{
  address={GSI Helmholtzzentrum f\"ur Schwerionenforschung GmbH, Planckstrasse 1, 64291 Darmstadt, Germany}
}

\begin{abstract}
We report on a novel scheme based on the chiral Lagrangian. It is used to analyze
pion-nucleon scattering, pion photoproduction, and nucleon Compton scattering.
Subthreshold partial-wave amplitudes
are calculated in chiral perturbation theory and analytically extrapolated
with constraints imposed by electromagnetic-gauge invariance, causality and unitarity. Experimental quantities  
are reproduced up to energies $\sqrt{s}\simeq 1300$ MeV in terms of the parameters relevant at order $Q^3$.
\end{abstract}

\maketitle


\section{Introduction}
Chiral perturbation theory has been extensively and successfully used as a
systematic tool for studying the low energy QCD dynamics. Particular
cases of pion-nucleon scattering, pion photoproduction and nucleon
Compton scattering were considered in \cite{Bernard:1995dp,Bernard:2007zu,Pascalutsa:2006up,
Bernard:1996gq,Fettes:1998ud,Fettes:2000xg,Beane:2004ra}.
The application of $\chi $PT is however limited to the near threshold region.

In this talk we report on a recently proposed scheme based on the chiral Lagrangian \cite{Gasparyan:2010xz}.
We describe  photon and pion scattering off the nucleon up to energies $\sqrt{s}\approx 1300$ MeV where the 
Lagrangian was truncated to order $Q^3$. An explicit isobar field is not considered. 
The physics of the isobar resonance enters our scheme by an infinite summation of higher order counter terms
in the chiral Lagrangian. The particular summation is performed in accordance with unitarity and causality.
The scheme is based on an analytic extrapolation of subthreshold scattering amplitudes that is controlled
by constraints set by electromagnetic-gauge invariance, causality and unitarity.
Unitarized scattering amplitudes are obtained which have left-hand cut structures in accordance with causality.
The latter are solutions of non-linear integral equations that are solved
by $N/D$ techniques.

\section{Description of the method}
The starting point of our method is the chiral Lagrangian involving pion,
nucleon and photon fields \cite{Fettes:1998ud,Bernard:2007zu}.
A strict chiral expansion of the amplitude to the order $Q^3$ includes
tree-level graphs, loop diagrams, and counter terms. Counter terms depend
on a few unknown parameters, which we fit to the empirical data on $\pi N$
elastic scattering and pion photoproduction. The Compton scattering amplitude
contains no counter terms to the order $Q^3$.

\begin{figure}[b]
 \includegraphics*[width=10cm]{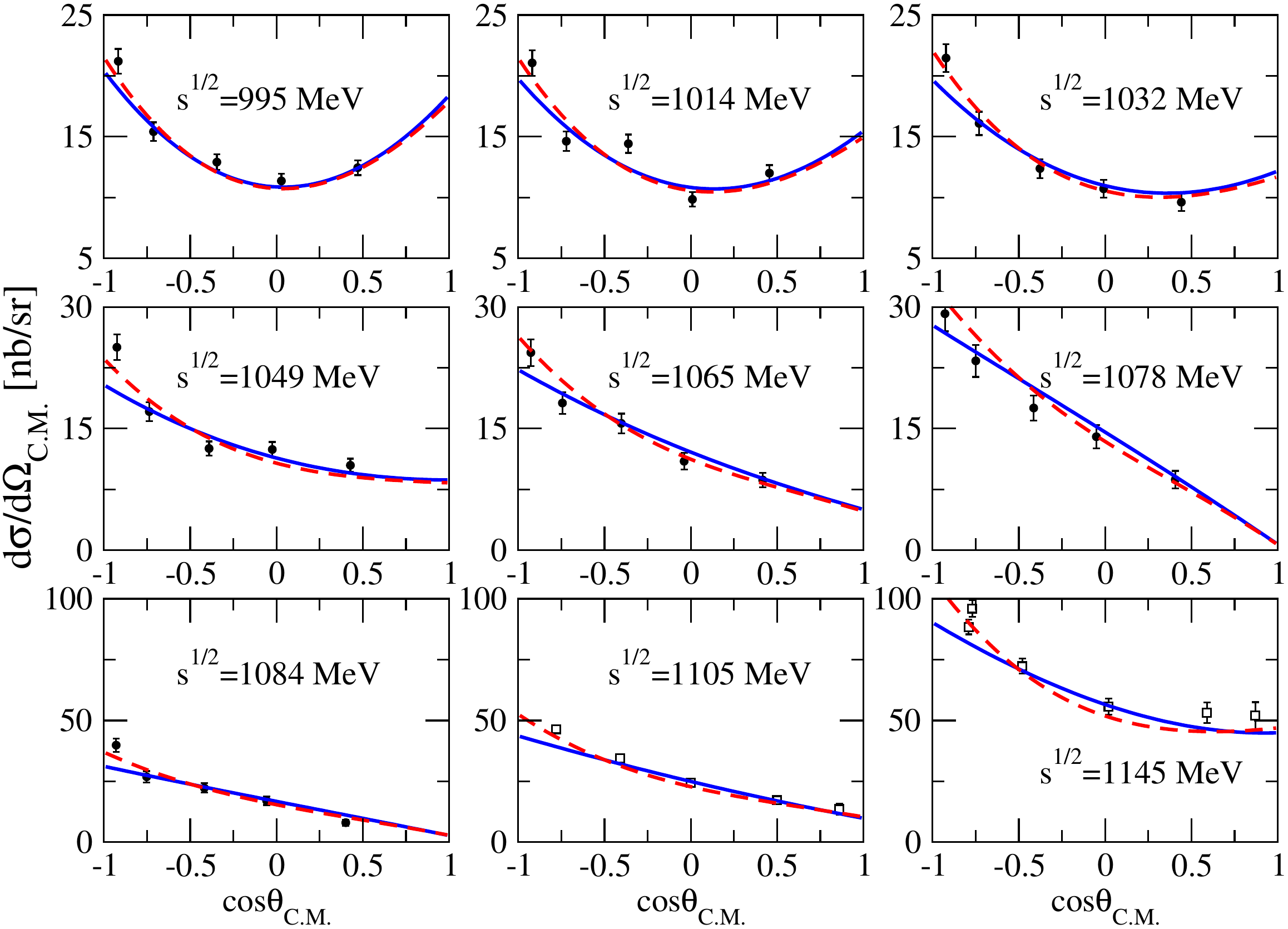}
\caption{Differential cross section for Compton scattering off the proton
as a function of the center-of-mass scattering angle. The data are from  \cite{OlmosdeLeon:2001zn} and
 \cite{Hallin:1993ft}. The solid lines follow from our partial-wave amplitudes with $J \leq 3/2$.
 The dashed lines show the effect of partial-wave contributions with $J> 3/2$.}
\label{fig:ComptondXS}
\end{figure}

Our approach is based on  partial-wave dispersion relations, for which the unitarity and causality constraints can be
combined in an efficient manner.
For a suitably chosen partial-wave amplitude with angular
momentum $J$, parity $P$ and channel quantum numbers $a,b$ (in our case $\pi N$ and $\gamma N$ channels)
we separate the right-hand cuts from the left-hand cuts
\begin{eqnarray}
 T_{ab}(\sqrt{s}\,)=U_{ab}(\sqrt{s}\,)+\sum_{c,d}\, \int_{\mu_{\rm thrs}}^{\infty}\frac{dw}{\pi}\,\frac{\sqrt{s}-\mu_M}{w-\mu_M}\,
\frac{T_{ac}(w)\,\rho_{cd}(w)\,T_{db}^*(w)}{w-\sqrt{s}-i\epsilon},
\label{def-non-linear}
\end{eqnarray}
where the generalized potential, $U_{ab}^{(JP)}(\sqrt{s}\,)$, contains left-hand cuts only, by definition.
The sum in  (\ref{def-non-linear}) runs over all possible
intermediate states, and $\rho^{(JP)}_{cd}(\sqrt{s}\,)$, is the phase-space matrix.
It is emphasized that the separation (\ref{def-non-linear}) is gauge invariant.
The non-linear integral equation (\ref{def-non-linear}) can be solved by means of $N/D$ techniques \cite{Chew:1960iv}.
The amplitude is represented as
\begin{eqnarray}
T_{ab}(\sqrt{s}\,)=\sum_c\,D^{-1}_{ac}(\sqrt{s}\,)\,N_{cb}(\sqrt{s}\,)\,,
\label{def-NoverD}
\end{eqnarray}
where $D_{ab}(\sqrt{s}\,)$ has no singularities  but the right-hand $s$-channel unitarity cuts.  The coupled-channel
unitarity condition is a consequence of the ansatz
\begin{eqnarray}
D_{ab}(\sqrt{s}\,)=\delta_{ab}-
\sum_{c}\int_{\mu_{ \rm thr}}^\infty \frac{dw}{\pi}\,\frac{\sqrt{s}-\mu_M}{w-\mu_M}\frac{N_{ac}(w)\,\rho_{cb}(w)}{w-\sqrt{s}}\,,
\label{def-D}
\end{eqnarray}
where $N_{ab}(\sqrt{s}\,)$ obeys the linear integral equation
\begin{eqnarray}
&&N_{ab}(\sqrt{s}\,)=U_{ab}(\sqrt{s}\,)
\nonumber\\
&& \qquad \qquad +\sum_{c,d} \int_{\mu_{\rm thr}}^\infty \frac{dw}{\pi}\,
\frac{\sqrt{s}-\mu_M}{w-\mu_M}\,\frac{N_{ac}(w)\,\rho_{cd}(w)\,[U_{db}(w)-U_{db}(\sqrt{s}\,)]}{w-\sqrt{s}}\,.
\label{Nequation0}
\end{eqnarray}
The key observation we exploit in this work is the fact that the solution of the nonlinear integral equation (\ref{def-non-linear})
requires the knowledge of the generalized potential for $\sqrt{s} > \mu_{\rm thr}$ only. Conformal mapping techniques may be used to
approximate the generalized potential in that domain efficiently based on the knowledge
of the generalized potential in a small subthreshold region.
Given a suitable conformal mapping, $\xi(\sqrt{s}\,)$ (transforming a domain $\Omega$ into a unit circle),
we seek to establish a representation of the generalized potential of the form
\begin{eqnarray}
&& U(\sqrt{s}\,)=U_{\rm inside}(\sqrt{s}\,)+U_{\rm outside}(\sqrt{s}\,)\,,
\label{expansion} \\
&& U_{\rm outside}(\sqrt{s}\,)=\sum\limits_{k=0}^\infty {U_k}\,\big[\xi(\sqrt{s}\,)\big]^k\,, \qquad
U_k=\frac{d^k U_{\rm outside}(\xi^{-1}(\xi))}{k!\,d\xi^k}\Big|_{\xi=0}\,,
\nonumber
\end{eqnarray}
where we allow for an explicit treatment of cut structures that are inside the domain $\Omega$.
Then the expanded potential in (\ref{expansion}) is regular enough for the integral
equation (\ref{def-non-linear}) to be well defined and amenable to a solution via (\ref{Nequation0}).

The decomposition (\ref{expansion}) is faithful for energies $\sqrt{s} \in \Omega$ only. Nevertheless,
the full scattering amplitude can be derived by a crossing transformation of the solution to (\ref{def-non-linear}). 
Such a construction is consistent with crossing symmetry if the solution to (\ref{def-non-linear}) and its crossing transformed form coincide in a matching region around $\sqrt{s } \simeq  \mu_M$. With our choice
\begin{eqnarray}
\mu_M = m_N \,,
\label{def-muM}
\end{eqnarray}
this is clearly the case (see also \cite{Lutz:2001yb}).

\clearpage

\section{Results}

\begin{figure}[t]
 \includegraphics*[width=10cm]{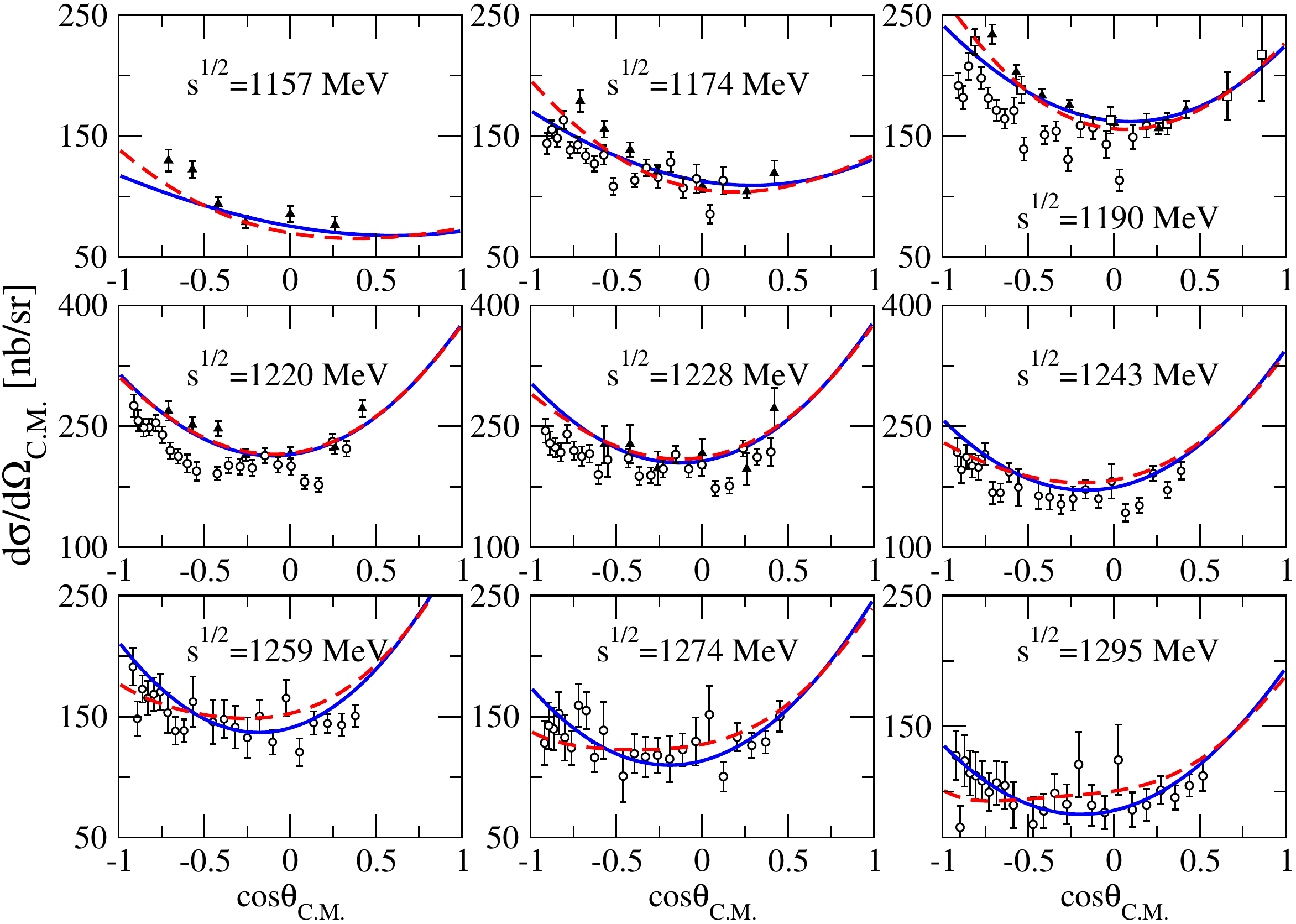}
\caption{Differential cross section for Compton scattering off the proton
as a function of the center-of-mass scattering angle.
The data are from   \cite{Hallin:1993ft},\cite{Blanpied:2001ae},
\cite{Wolf:2001ha} and the  lines are as in Fig. \ref{fig:ComptondXS}.}
\label{fig:ComptondXS2}
\end{figure}

\begin{figure}[b]
 \includegraphics*[width=10cm]{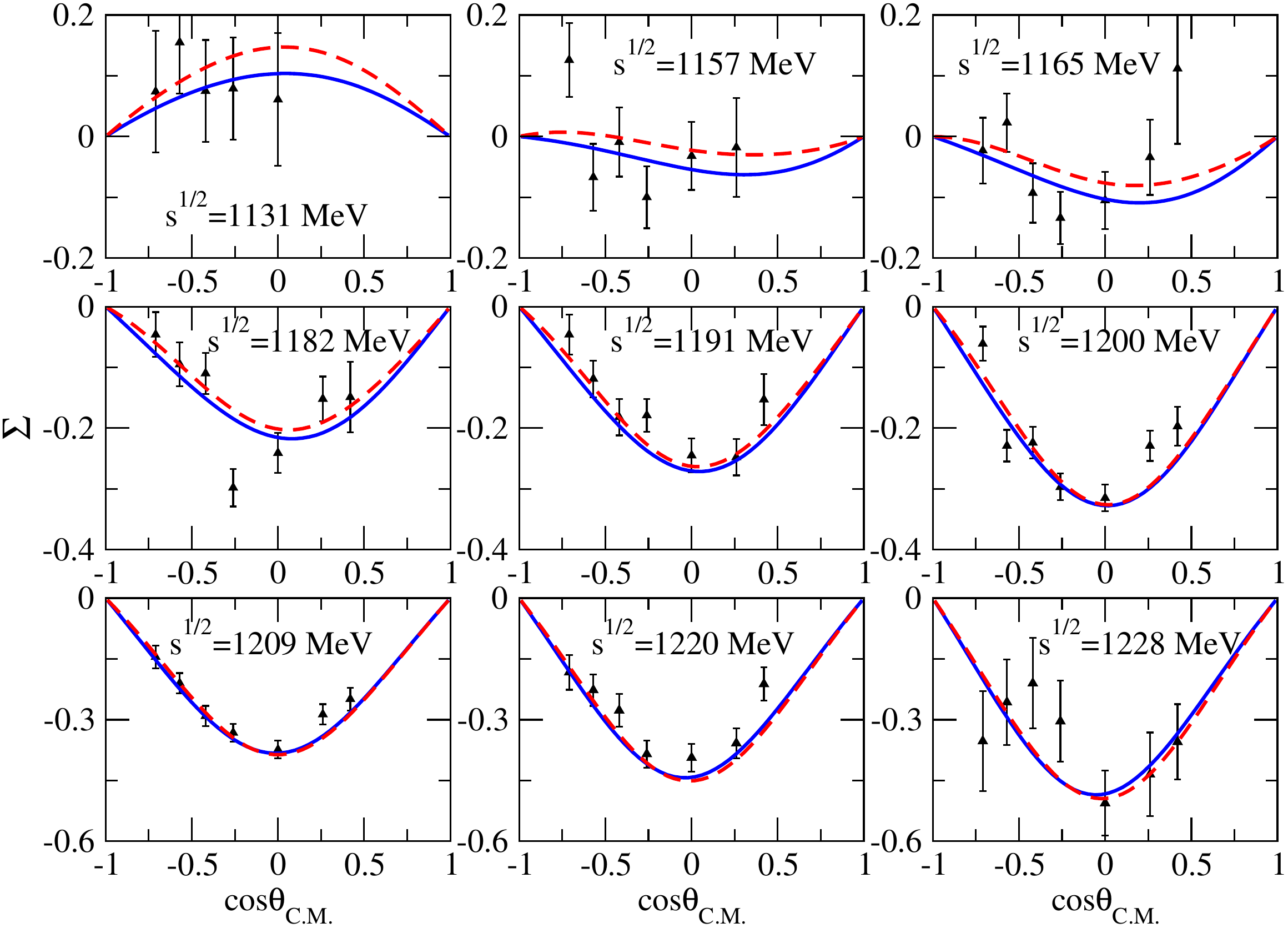}
\caption{Beam asymmetry for Compton scattering off the proton as a function of the center-of-mass scattering angle.
The data are from \cite{Blanpied:2001ae} and the lines  are as in Fig. \ref{fig:ComptondXS}.}
\label{fig:ComptonSigma}
\end{figure}

\begin{figure}[t]
 \includegraphics*[width=10cm]{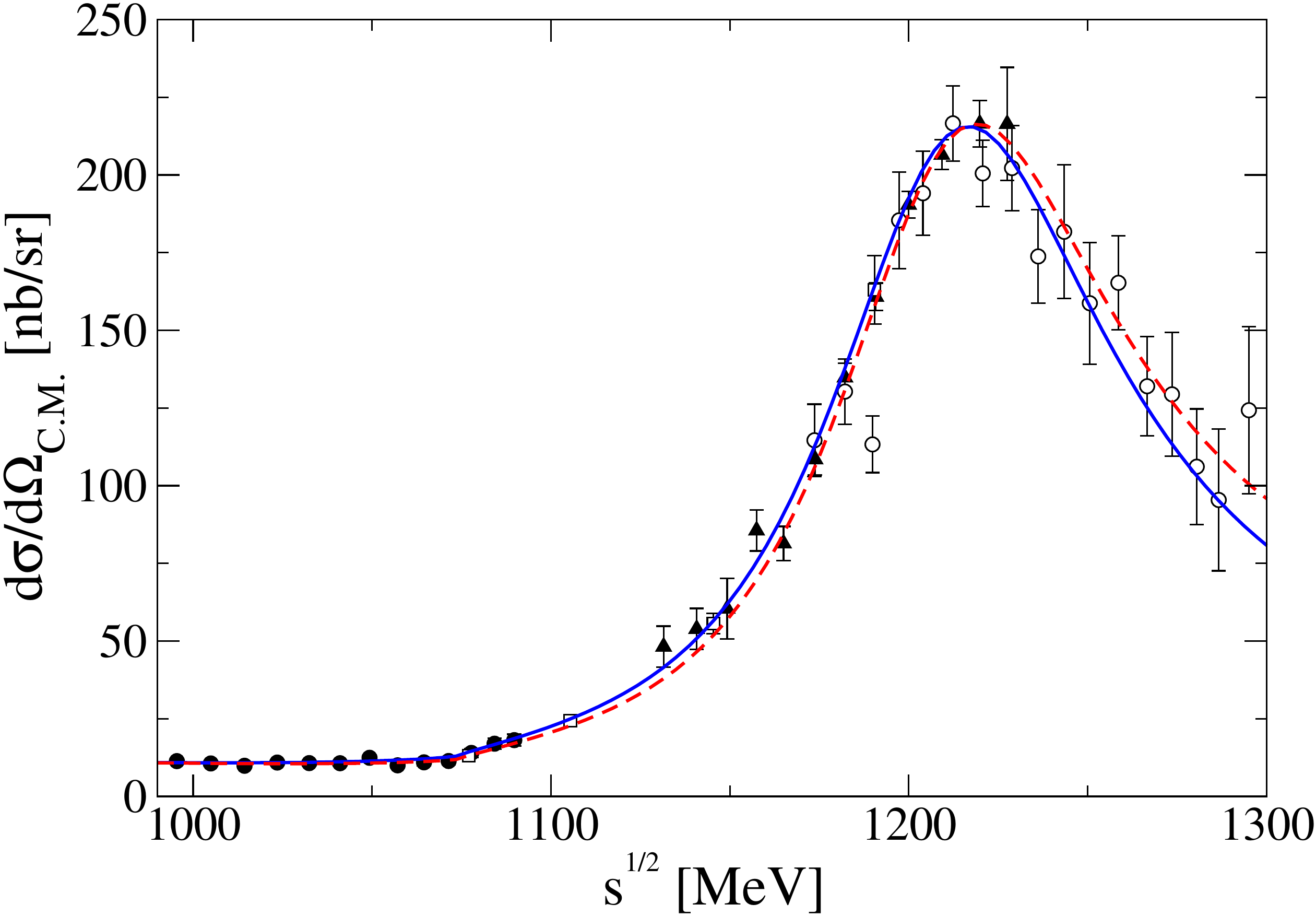}
\caption{Compton scattering off the proton as a function of energy at scattering angle $\theta=90^\circ$.
The data are from  \cite{OlmosdeLeon:2001zn},
 \cite{Hallin:1993ft}, \cite{Blanpied:2001ae},
\cite{Wolf:2001ha} and the  lines are as in Fig. \ref{fig:ComptondXS}.
The experimental values correspond to scattering angles closest to $\theta=90^\circ$ taken in the interval
$86^\circ<\theta<94^\circ$.}
\label{fig:Compton90}
\end{figure}

We determined the counter terms of the $\pi N$ sector as to reproduce the empirical s- and p-wave  $\pi N$
elastic phase shifts. On this example we demonstrated reasonable convergence
properties of the method with respect to the chiral expansion parameter $Q$ up to total energies of about $1300$ MeV. 
Additional parameters are relevant for pion photoproduction. Having adjusted those parameters we are able to
reproduce a wide variety of pion photoproduction data, such as differential cross sections and spin observables for different
reaction channels with good accuracy. Most predictive is proton Compton scattering, which involves no additional free parameter. 
Our results for the differential cross section and beam asymmetry are presented in Figs.~\ref{fig:ComptondXS}-\ref{fig:ComptonSigma}
against empirical data. We find agreement with the data taking into account some discrepancy of the
different data sets.  The photon threshold region, the pion-production threshold region, and the isobar region are equally well
reproduced. This is nicely illustrated by Fig. \ref{fig:Compton90}, which shows the energy dependence of the cross
section at fixed scattering angle  $\theta \simeq 90^\circ$.

\section{Summary}

We reviewed recent progress on a unified description of pion and photon-nucleon scattering up to and
beyond the isobar resonance region. A novel scheme that combines causality, unitarity and gauge invariance
and that is based on the chiral Lagrangian was discussed. We focused on results obtained for proton Compton scattering.

\bibliographystyle{aipproc}   

\bibliography{sample}

\end{document}